\documentclass{article}
\usepackage{ijcai22}

\usepackage{times}
\usepackage{soul}
\usepackage{url}
\usepackage[hidelinks]{hyperref}
\usepackage[utf8]{inputenc}
\usepackage[small]{caption}
\usepackage{graphicx}
\usepackage{amsmath}
\usepackage{amsthm}
\usepackage{amssymb}
\usepackage{booktabs}
\usepackage{algorithm}
\usepackage{algorithmic}
\usepackage{verbatim}
\usepackage{multirow}
\usepackage{multicol}
\usepackage{amssymb}
\usepackage{subfigure}
\usepackage{bbding}
\usepackage{amssymb}
\usepackage{color}

\title{{Taylor, Can You Hear Me Now?} A Taylor-Unfolding Framework for Monaural Speech Enhancement}
\author{
Andong Li$^{1,2}$
\and
Shan You$^{3}$\and
Guochen Yu$^{1, 4}$\and
Chengshi Zheng$^{1,2}$\thanks{Contact Author}\and
Xiaodong Li$^{1,2}$
\affiliations
$^1$Institute of Acoustics, Chinese Academy of Sciences\\
$^2$University of Chinese Academy of Sciences\\
$^3$SenseTime Research \  $^4$Communication University of China
\emails 
\{liandong, cszheng, lxd\}@mail.ioa.ac.cn, youshan@sensetime.com, yuguochen@cuc.edu.cn
}

\begin{document}
\maketitle
\begin{abstract}
While the deep learning techniques promote the rapid development of the speech enhancement (SE) community, most schemes only pursue the performance in a black-box manner and lack adequate model interpretability. Inspired by Taylor's approximation theory, we propose an interpretable decoupling-style SE framework, which disentangles the complex spectrum recovery into two separate optimization problems \emph{i.e.}, magnitude and complex residual estimation. Specifically, serving as the 0th-order term in Taylor's series, a filter network is delicately devised to suppress the noise component only in the magnitude domain and obtain a coarse spectrum. To refine the phase distribution, we estimate the sparse complex residual, which is defined as the difference between target and coarse spectra, and measures the phase gap. In this study, we formulate the residual component as the combination of various high-order Taylor terms and propose a lightweight trainable module to replace the complicated derivative operator between adjacent terms. Finally, following Taylor's formula, we can reconstruct the target spectrum by the superimposition between 0th-order and high-order terms. Experimental results on two benchmark datasets show that our framework achieves state-of-the-art performance over previous competing baselines in various evaluation metrics. The source code is available at \href{https://github.com/Andong-Li-speech/TaylorSENet}{github.com/Andong-Li-speech/TaylorSENet}.
\end{abstract}
\section{Introduction}
\label{sec:introduction}
As a consequence of the COVID-19 pandemic, people and organizations have become increasingly dependent on remote communication techniques to stay connected and conduct business routines. It is thus imperative to demand high-quality speech when background noise and room reverberation exist. As a resolution, monaural speech enhancement (SE) aims to extract the target speech from the noisy mixture when only the single-channel recording is available. Recently, the advent of deep neural networks (DNNs) has significantly promoted the performance of SE algorithms, which can be roughly categorized into two streams, namely in the time-domain~{\cite{pascual2017segan,luo2019conv}} and time-frequency (T-F)-domain~{\cite{yin2020phasen,tang2021joint}}. As speech and noise patterns tend to be more distinguishable after the short-time Fourier transform (STFT), the latter still dominates the mainstream.
\begin{figure}[t]
	\centering
	\centerline{\includegraphics[width=1.00\columnwidth]{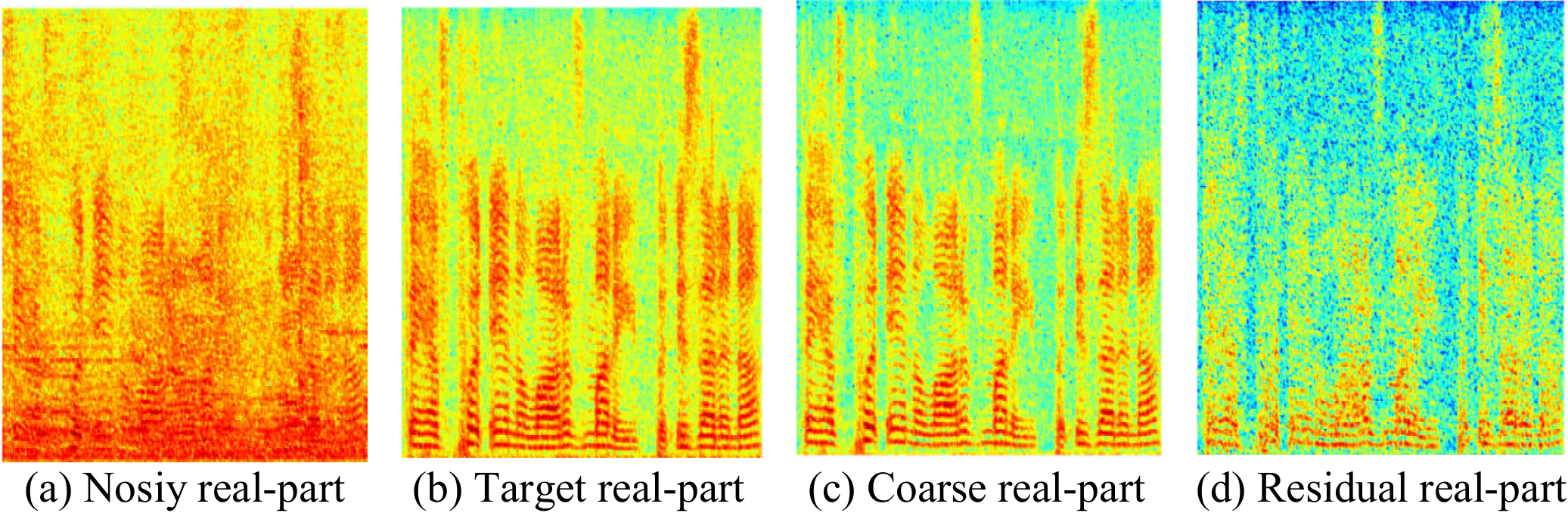}}
	\caption{Example of the decoupling target, and only the real part is visualized for illustration convenience.}
	\label{fig:example}
	\vspace{-0.5cm}
\end{figure}
Previous works simply estimated the magnitude of the spectrum and left the noisy phase unaltered, but they would inevitably incur heavy performance restrictions. To address this problem, it is necessary to consider the joint optimization of magnitude and phase. In~{\cite{yin2020phasen}}, a dual-branch network was designed to separately model the magnitude and cosine representations of phase. However, due to the nonstructural characteristic of phase, the phase-branch tends to be sensitive to nonlinear operations. Another typical strategy is to couple the magnitude and phase into Cartesian coordinates and construct real and imaginary (RI) pairs. By virtue of complex spectral mapping (CSM)~{\cite{tan2020learning}} or complex ratio mask (CRM)~{\cite{williamson2015complex}}, both the magnitude and phase can be implicitly recovered. However, such target entanglement will cause the compensation effect~{\cite{wang2021compensation}}, \emph{i.e.}, magnitude distortion is inevitably sacrificed to compensate for the phase prediction accuracy, especially under low signal-to-noise ratios (SNRs).

More recently, a decoupling mapping procedure~{\cite{li2021two}} was proposed to decouple the complex spectrum estimation into two separate steps. In the first step, only the magnitude prior is estimated, which is coupled with the noisy phase to obtain a coarse complex spectrum. Afterward, with residual learning, another network is utilized to estimate the residual component with sparse distribution in the complex domain, which measures the gap between noisy and target phases. Different from previous literature, it endows the separate optimization space toward the magnitude and phase, and therefore, alleviates the compensation effect. Figure~{\ref{fig:example}}(c)-(d) visualize the coarse and residual spectra as an example.

In this paper, we rethink the spectrum decoupling and formulate it as an approximation problem \emph{w.r.t.} input neighborhood space. In other words, \textit{if we can access the complex residual and repair the phase in advance, then we can perfectly approximate the clean spectrum from magnitude estimation theoretically}. This process can be expressed as $S = \mathcal{F}\left(X + \delta\right)$, where $\left\{S, X, \delta\right\}$ denote clean, mixture, and residual components, respectively, and $\mathcal{F}$ is the magnitude estimation function. Based on this new formulation, it is intuitive to leverage Taylor's approximation \emph{w.r.t.} $X$ to estimate the function representation at $X + \delta$. However, in practical implementation, the residual prior is usually inaccessible. In this regard, we propose a new framework called \textbf{TaylorSENet} to explicitly model the Taylor's approximation by formulating the main term and derivative term as learnable modules. Concretely, the 0th-order module is to consider the magnitude of the spectrum while the high-order modules are concerned with complex residual estimation. In this way, the estimation of the complex spectrum can be obtained by the superimposition of the 0th-order non-derivative and multiple high-order derivative terms. Different from previous SE models in a black-box manner, the proposed framework provides each module with better interpretability. To the best of our knowledge, this is the first time to cast the complex spectrum recovery as Taylor's approximation problem in the speech front-end task. Our contributions can be summarized as three-fold:
\begin{itemize}
	\item We rethink the decoupling-style SE algorithm and abstract it as Taylor's approximation problem.
	\item We propose an end-to-end framework to simulate the 0th-order and high-order items of Taylor-unfolding.
	\item We conduct comprehensive experiments and the results show that our system achieves state-of-the-art performance among two benchmarks.
\end{itemize}
\vspace{-0.3cm}
\section{Related Work}
\label{sec:related-work}
\paragraph{T-F Domain Methods.}
Before entering the deep learning era, traditional denoising algorithms were applied in the T-F domain as Fourier theory provides a feasible feature representation. Typical methods include spectral subtraction~{\cite{boll1979suppression}}, Wiener filtering~{\cite{scalart1996speech}}, and statistical-based methods~{\cite{ephraim1984speech}}. After the proliferation of DNNs, the denoising task is formulated into a supervised learning problem and can be trained to grasp the latent mapping relations between noisy features and clean targets. For a long time, only the magnitude is considered as the phase distribution is nonstructured and is thus difficult to predict. More recently, increasing evidence shows that phase also plays a pivotal role in perception quality improvement~{\cite{paliwal2011importance}}.

For phase-aware SE approaches, in GCRN~{\cite{tan2020learning}}, a UNet-style network was utilized to estimate both real and imaginary (RI) parts of the spectrum. As a modification, DCCRN~{\cite{hu2020dccrn}} devised a complex-valued UNet to adapt the complex correlation between RI. {\cite{yin2020phasen}} proposed a dual-branch network to model the magnitude filter and cosine representations of the phase, respectively. CTSNet~{\cite{li2021two}} proposed a two-stage mapping regime, where the magnitude was first estimated as the prior to further facilitate the subsequent phase recovery. 
\paragraph{Time Domain Methods.}
Thanks to the development of DNNs, time-domain-based methods have gained prosperity more recently. A typical option is to directly learn the sample distribution. In DDAEC~{\cite{pandey2020densely}}, the 1-D waveform was first enframed into 2-D format and then passed through a UNet structure with layer-wise dense-nets. SEGAN~{\cite{pascual2017segan}} adopted the generative adversarial network (GAN) to predict the waveform directly. Another strategy is to use a pair of learnable encoder and decoder to convert the waveform samples into latent space and then devise a separate module to distinguish between different sources. Representative works include Conv-TasNet~{\cite{luo2019conv}} and DPRNN~{\cite{luo2020dual}}.
\paragraph{Multi-stage Learning.}
Due to the lack of prior information, the performance of a single-stage SE pipeline is heavily limited in complicated acoustic scenarios. In contrast, in a multi-stage pipeline, the original mapping problem is usually decomposed into several separate subtasks and enables progressive learning. Besides, the previous estimation also serves as the latent prior to guide the subsequent learning process. {\cite{westhausen2020dual}} combined the complementarity of T-F and latent domain and proposed a stacked dual signal transformation network. {\cite{hao2021fullsubnet}} rethinked spectrum recovery from the dual optimization of subband and fullband and proposed a two-stage approach to capture both local and global spectral contexts. Based on the glance and gaze behavior of humans in visual perception, {\cite{li2022glance}} proposed to stack multiples glance-gaze modules to reconstruct the spectrum collaboratively.
\begin{figure*}[t]
	\centering
	\centerline{\includegraphics[width=1.6\columnwidth]{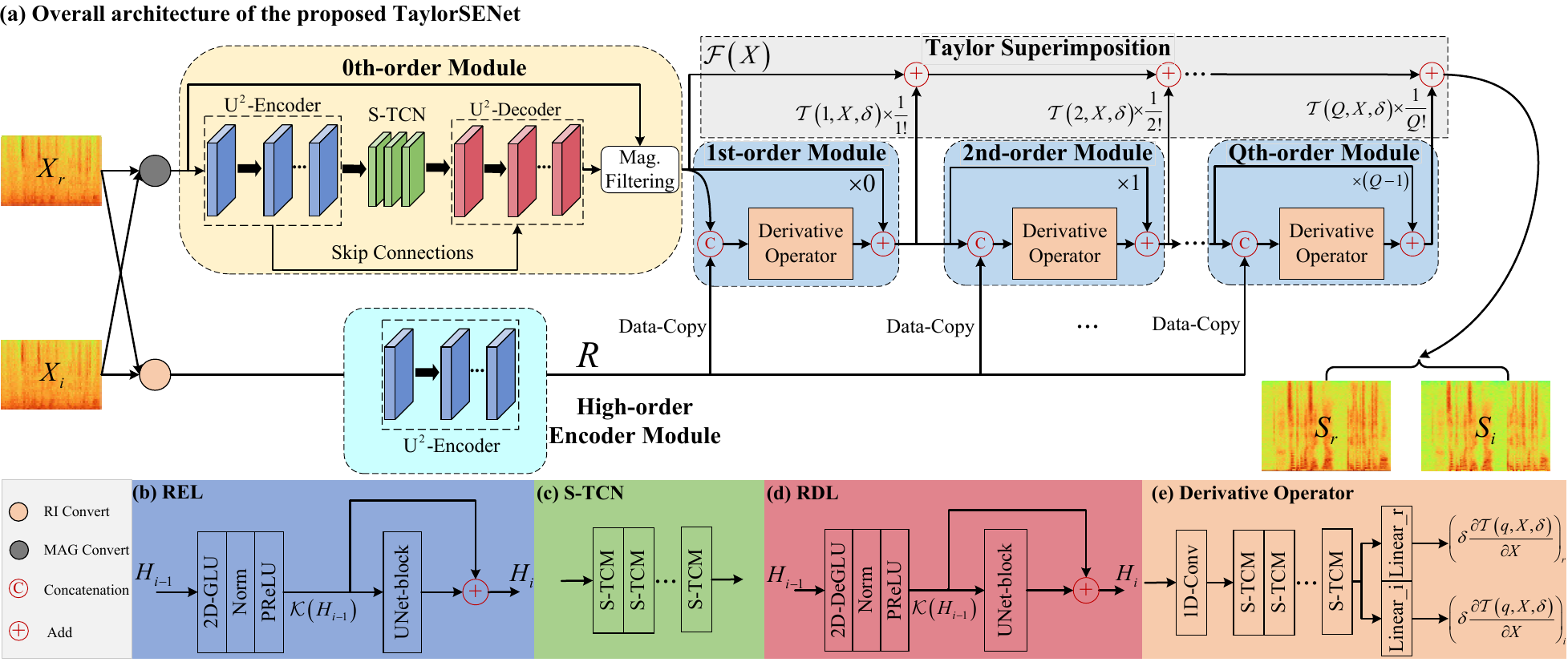}}
	\caption{An overview of the proposed TaylorSENet. (a) Network diagram of TaylorSENet, which involves the superimposition of multiple-order terms. The 0th-order module adopts the magnitude filtering. The calculation of high-order terms follows the recursive formula in Eqn.{(\ref{eqn13})}. (b) Detail of the recalibration encoding layer (REL) in the U$^{2}$ encoder. (c) Detail of the squeezed-TCN (S-TCN). (d) Detail of the recalibration decoding layer (RDL) in the U$^{2}$ decoder. (e) Detail of the derivative operator.}
	\label{fig:framework}
	\vspace{-0.5cm}
\end{figure*}
\vspace{-0.12cm}
\section{Methodology}
\label{methodology}
\subsection{From Decoupling to Taylor's Series Expansion}
Given STFT, the mixture signal in the T-F domain can be formulated as:
\begin{equation}
\label{eqn1}
X_{k, l} = S_{k, l} + N_{k, l},
\end{equation}
where $\left\{X_{k, l}, S_{k, l}, N_{k, l}\right\}\in\mathbb{C}$ denote complex-valued noisy, clean, and noise signals in the frequency bin of $k\in\left\{1,\cdots, K\right\}$ and time index of $l\in\left\{1,\cdots,L\right\}$. For brevity, we omit the subscripts $\left\{k, l\right\}$ if no confusion arises. The aim of SE is to design an operator to extract the target speech from the noisy mixture, \emph{i.e.}, 
\begin{equation}
\label{eqn2}
S = \mathcal{F}_{1}\left(X\right),
\end{equation}
where $\mathcal{F}_{1}$ denotes the estimation function. Although various networks can be employed to accomplish this process, these methods usually encapsulate the whole recovery process as a black-box and thus have weak interpretability in the intermediate stages~{\cite{tan2020learning,hu2020dccrn}}. Recently, a decoupling-style forward pipeline was proposed in {\cite{li2021two}} and given as
\begin{align}
\label{eqn3}
&\left|S\right| = \mathcal{F}_{mag}\left(\left|X\right|\right),\\
&S = \left|S\right|e^{j\theta_{X}} + \mathcal{F}_{com}\left(S, X\right),
\end{align}
where $\mathcal{F}_{mag}$ and $\mathcal{F}_{com}$ are the mapping functions for magnitude and residual estimation, respectively, and $\theta_{X}$ denotes the noisy phase. The above procedure decomposes the entanglement of magnitude and phase by step-wise optimization, and can be summarized into two core operations: 
\begin{itemize}
	\item \textbf{op1}: suppress the noise in the magnitude domain while ignoring the phase term to obtain a coarse estimation.
	\item \textbf{op2}: estimate the complex residual while fixing the magnitude term to refine the target spectrum.
\end{itemize}

In traditional SE algorithms, \textbf{op1} can be fulfilled by either noise subtraction~{\cite{boll1979suppression}} or noise filtering~{\cite{ephraim1984speech}}. For both two techniques, the aforementioned procedure can be respectively expressed as
\begin{align}
\label{eqn4}
S &= \left(\left|X\right| - \left|N\right|\right)e^{j\theta_{X}} + \delta_{1} = X - \widetilde{N}_{1} + \delta_{1},\\
S &= M\left|X\right|e^{j\theta_{X}} + \delta_{2} = X - \widetilde{N}_{2} + \delta_{2},
\end{align}
where $\left\{\widetilde{N}_{1} = \left|N\right|e^{j\theta_{X}}, \widetilde{N}_{2} = (1 - M)\left|X\right|e^{j\theta_{X}}\right\}$ respectively denote the estimated noise components in traditional noise subtraction and noise filtering algorithms, and $M$ denotes the spectral filter gain. $\left\{\delta_{1}, \delta_{2}\right\}$ denote the corresponding phase residual term.

Comparing Eqn.(5) and Eqn.(6), we can find that they have a similar format, which is intuitive since \textbf{op1} essentially encourages an accurate noise power spectral density (NPSD) estimation and then subtracts it from the mixture. Furthermore, denoting $\mathcal{F}\left(X\right) = X - \widetilde{N}$, and $X:=X+\delta$, Eqns.(5)-(6) can thus be abstracted into a more general case:
\begin{equation}
\label{eqn6}
S = \mathcal{F}\left(X + \delta\right).
\end{equation}

Eqn.(\ref{eqn6}) implies that if we can access the residual term $\delta$ and add it into input $X$ in advance, then we are able to perfectly recover the spectrum by magnitude estimation theoretically. However, in practical scenarios, we usually can not access the residual prior $\delta$. Therefore, to resolve the above generalized function, we expand Eqn.(\ref{eqn6}) with an infinite Taylor's series expansion at $X$, given as
\begin{align}
\label{eqn7}
\resizebox{0.91\linewidth}{!}{$
S = \mathcal{F} \left(X\right) + \frac{1}{1!}\frac{\partial\mathcal{F}\left(X\right)}{\partial X}\delta+...+ \frac{1}{q!}\frac{\partial^{q}\mathcal{F}\left(X\right)}{\partial^{q}X}\delta^{q}+...
$}
\end{align}
which can be simplified as
\begin{equation}
\label{eqn8}
S = \underbrace{\mathcal{F}\left(X\right)}_{\text{0th-order term}} + \underbrace{\sum_{q=1}^{+\infty}\frac{1}{q!}\frac{\partial^{q}\mathcal{F}\left(X\right)}{\partial^{q} X}\delta^{q}}_{\text{high-order terms}}.
\end{equation}

As such, we provide a novel perspective of complex spectrum recovery from Taylor's approximation theory. In detail, the 0th-order term is grounded with regard to the estimation of spectral magnitude, and the high-order terms attempt to approximate the distribution of the residual component.
\vspace{-3pt}
\subsection{Taylor-Unfolding Framework}
\label{sec:taylor-unfolding-framework}
To adapt the formulation of Taylor's series expansion to the network, it is necessary to derive the correlation between adjacent derivative terms. For practical implementation, we truncate the number of orders in the derivative part into $Q$ and neglect higher-order parts. In light of~{\cite{fu2021unfolding}}, we first define the $q$th order derivative term as $\mathcal{T}\left(q, X, \delta\right)$
\begin{equation}
\vspace{-0.05cm}
\label{eqn9}
\mathcal{T}\left(q, X, \delta\right) = \frac{\partial^{q}\mathcal{F}\left(X\right)}{\partial^{q} X}\delta^{q},
\vspace{-0.05cm}
\end{equation}
where the factorial term is dropped for derivation convenience. To investigate the correlation between $q$th order $\mathcal{T}\left(q, X, \delta\right)$ and $(q+1)$th order $\mathcal{T}\left(q+1, X, \delta\right)$, we differentiate $\mathcal{T}\left(q, X, \delta\right)$ \emph{w.r.t.} $X$
\begin{equation}
\label{eqn10}
\resizebox{0.90\linewidth}{!}{$
\frac{\partial\mathcal{T}\left(q,X,\delta\right)}{\partial X} = \frac{\partial}{\partial X}\left(\frac{\partial^{q}\mathcal{F}\left(X\right)}{\partial^{q} X}\right)\delta^{q} + \frac{\partial^{q}\mathcal{F}\left(X\right)}{\partial^{q} X}\frac{\partial}{\partial X}\delta^{q}.
$}
\end{equation}

Considering
\begin{equation}
\label{eqn11}
\frac{\partial}{\partial X}\left(\frac{\partial^{q}\mathcal{F}\left(X\right)}{\partial^{q} X}\right) = \frac{\partial^{q+1}\mathcal{F}\left(X\right)}{\partial^{q+1} X},
\vspace{-0.05cm}
\end{equation}
\begin{equation}
\label{eqn12}
\frac{\partial}{\partial X}\delta^{q} = \frac{\partial\delta^{q}}{\partial \delta}\frac{\partial\delta}{\partial X} = -q\delta^{q-1}.
\end{equation}

Substituting Eqns.(\ref{eqn11})-(\ref{eqn12}) into Eqn.(\ref{eqn10}) and multiplying $\delta$ on both sides, we can derive the following recursive formula between $\mathcal{T}\left(q+1, X, \delta\right)$ and $\mathcal{T}\left(q, X, \delta\right)$:
\begin{gather}
\label{eqn13}
\vspace{-0.1cm}
\mathcal{T}\left(q+1, X, \delta\right) = q\mathcal{T}\left(q,X,\delta\right) + \delta\frac{\partial\mathcal{T}\left(q,X,\delta\right)}{\partial X}.
\vspace{-0.1cm}
\end{gather}

Generally speaking, it is quite difficult to access the derivative operator $\delta\frac{\partial\mathcal{T}\left(q,X,\delta\right)}{\partial X}$. To this end, we design a trainable network module, notated as $\mathcal{P}\left(q,X,\delta\right)$, to replace the complicated operator. Note that as the network weights are purely learned from the training data, it does not necessarily follow a strict mathematical definition of derivation, but we empirically find that it indeed involves the residual estimation. 
\vspace{-0.12cm}
\subsection{Framework Structure}
\label{network-structure}
\subsubsection{Forward Stream}
\label{forward-stream}
We instantiate the proposed Taylor-unfolding framework, as shown in Figure~{\ref{fig:framework}}(a). It mainly comprises two parts, namely the 0th-order module and multiple high-order modules. According to our previous formulation, the 0th-order module targets at magnitude estimation. To this end, we first convert the input RI into magnitude, \emph{i.e.}, $\lvert X\rvert = \sqrt{X_{r}^{2} + X_{i}^{2}}$, and then we sent it to the 0th-order module to obtain the output gain $M$ with range $\left(0, 1\right)$ for noise filtering in the magnitude domain, as shown in Eqn.{(15)}. To model high-order terms, we employ the high-order encoder to directly extract the patterns from the RI input, and the output feature map $R$ is then concatenated with the output from the last high-order module as the input of the next module for high-order term update, as shown in Eqns.(16)-(17). This operation is implemented recursively. Note that after magnitude filtering, we couple the filtered spectral magnitude with the noisy phase $\theta_{X}$ to yield the coarse complex spectrum. After all the terms are obtained, following Taylor's formula, we superimpose all of them to reconstruct the target spectrum. In a nutshell, the overall forward stream is formulated as:
\begin{align}
\mathcal{F}\left(X\right) &= M\left| X \right|e^{j\theta_{X}},\\
\mathcal{P}\left(q, X, \delta\right) &= \mathcal{G}\left(Concat\left(\mathcal{T}\left(q, X,\delta\right), R\right)\right), \\
\mathcal{T}\left(q + 1, X, \delta\right) &= q \odot\mathcal{T}\left(q, X, \delta\right) + \mathcal{P}\left(q, X, \delta\right), \\
S &= \mathcal{F}\left(X\right) + \sum_{q=1}^{Q}\frac{1}{q!}\mathcal{T}\left(q, X, \delta\right),
\vspace{-0.2cm}
\end{align}
where $q\in\left\{1,\cdots,Q\right\}$ is the order index, and $\mathcal{G}$ denotes the mapping function of derivative operator in the high-order module. $\odot$ denotes the Hadamard product.
\vspace{-0.05cm}
\subsubsection{0th-Order Module}
\label{Zero-order-module}
In the 0th-order module, we adopt a classical UNet-style encoder-decoder structure, which has been widely utilized in the SE task~{\cite{tan2020learning,li2021two}}. The encoder is to gradually decrease the feature size with consecutive downsampling operations while extracting the spectral features. In contrast, the decoder has a mirror structure and attempts to recover the original spectral size with deconvolution layers. Nonetheless, high-level semantic information is usually embedded in the various-length frame correlations and naive convolution operations often can not capture such complicated multiscale information. Inspired by the success of U$^{2}$-Net in the salient object detection field~{\cite{qin2020u2}}, we adapt the U$^{2}$-Encoder and U$^{2}$-decoder with multiple recalibration encoding/decoding layers (REL/RDLs) herein, as shown in Figure~{\ref{fig:framework}}(b)(d). Taking REL as an example, each 2D-gated linear unit (GLU)~{\cite{dauphin2017language}} is followed by instance normalization (IN), and PReLU. Then a UNet-block is inserted with the residual connection, which takes the UNet-style structure except that the layer depth dynamically varies with the current input size. The process can be expressed as:
\begin{align}\label{eqn16}
\mathcal{K}\left(H_{i-1}\right) &= \text{PReLU}\left(\text{Norm}\left(\text{2D-GLU}\left(H_{i-1}\right)\right)\right),\\
H_{i} &= \mathcal{K}\left(H_{i-1}\right) + \text{UNet-block}\left(\mathcal{K}\left(H_{i-1}\right)\right),
\end{align}
where $H_{i}$ denotes the input feature map of the $i$-th REL. The rationale is two-fold. First, by further feature downsamling-upsampling operations, different scale information can be effectively grasped. Besides, as the feature map close to the input tends to be rather noisy, we can recalibrate the feature map and preserve the target information.

To establish long-term relations between adjacent frames, we stack multiple temporal convolution networks (TCNs)~{\cite{luo2019conv}}, which comprise multiple temporal convolution modules (TCMs) with increasing dilations in the time axis. Besides, to decrease the computational footprint, we adopt a squeezed version of TCN (dubbed S-TCN)~{\cite{li2021two}}, as shown in Figure~{\ref{fig:framework}}(c), where the squeezed TCM (S-TCM) is leveraged for more compact channels. We also investigate other advanced structures for 0th-order module such as transformers~{\cite{vaswani2017attention}} and conformers~{\cite{gulati2020conformer}} in experimental ablation studies (see Section~{\ref{sec:ablation-study}).
\vspace{-0.1cm}
\subsubsection{High-Order Module}
\label{high-order-module}
For effective network training, we model the complicated derivative operator with a trainable network module, whose internal structure is presented in Figure~{\ref{fig:framework}}(e). As Eqn.{(\ref{eqn13})} indicates, the operator involves both the last-order term and input $X$, so we utilize both the encoded feature $R$ from noisy input $X$ and the estimation from the last-order term $\mathcal{T}\left(q-1,X,\delta\right)$ as the input and send to a 1D-Conv. Several S-TCMs are employed as the modeling unit, and we can generate $\delta\frac{\partial\mathcal{T}\left(q, X, \delta\right)}{\partial X}$ with linear transformation.
Note that our framework also applies to other more advanced network structures and are expected to achieve even better performance, which we leave as future work. Besides, as the derivative operator should be parameter-invariant theoretically, we also investigate the case when the parameters are shared among different derivative modules (see Section~{\ref{sec:ablation-study}}).
\section{Experiments}
\label{experiments}
\subsection{Datasets}
\paragraph{WSJ0-SI84.} It consists of 7138 utterances by 83 speakers (42 males and 41 females)~{\cite{paul1992design}}. We randomly select 5428 and 957 clips for training and validation, and another 150 clips by untrained speakers are used for testing. To generate noisy-clean pairs, we sample around 20,000 noises from the DNS-Challenge noise set~{\cite{reddy2020interspeech}} and the training SNRs are sampled from [-5\rm{dB}, 0\rm{dB}]. As a result, approximately 300 hours pairs are created for training. For model evaluation, three untrained challenging noises are selected, namely babble, factory1, and cafeteria. Three testing SNRs are set, \emph{i.e.}, $\left\{-3\rm{dB}, 0\rm{dB}, 3\rm{dB}\right\}$, and 150 pairs are created for each case.
\paragraph{DNS-Challenge.} The Interspeech 2020 DNS-Challenge corpus covers over 500 hours of clean clips by 2150 speakers and over 180 hours of noise clips~{\cite{reddy2020interspeech}}. For model evaluation, it provides a non-blind validation set with two categories, namely with and without reverberation, and each includes 150 noisy-clean pairs. Following the scripts provided by the organizer, we generate around 3000 hours of noisy-clean pairs for training and the SNRs randomly range from -5\rm{dB} to 15\rm{dB}.
\renewcommand\arraystretch{0.90}
\begin{table}[t]
	\Huge
	\centering
	\resizebox{\columnwidth}{!}{
		\begin{tabular}{c|ccccccc}
			\specialrule{0.1em}{0.25pt}{0.25pt}
			\multirow{2}*{Entry} &Param.  &Is- &\multirow{2}*{$Q$} &Zero- &\multirow{2}*{PESQ$\uparrow$} &\multirow{2}*{ESTOI(\%)$\uparrow$} &\multirow{2}*{SISNR(dB)$\uparrow$}\\
			&(M) &shared & &type & & &\\
			\specialrule{0.1em}{0.25pt}{0.25pt}
			1a &\textbf{2.19} &$\times$ &0 &U$^{2}$ &2.63 &69.66 &8.48\\
			1b &3.76 &$\times$ &1 &U$^{2}$ &2.76 &74.38 &10.25\\
			1c &4.58 &$\times$ &2 &U$^{2}$ &2.80 &75.35 &10.54\\
			1d &5.40 &$\times$ &3 &U$^{2}$ &\textbf{2.81} &76.05 &10.79\\
			1e &6.22 &$\times$ &4 &U$^{2}$ &2.80 &75.66 &10.66\\
			1f &7.04 &$\times$ &5 &U$^{2}$ &2.80 &\textbf{76.09} &\textbf{10.85}\\
			\specialrule{0.1em}{0.25pt}{0.25pt}
			2c &3.76 &\checkmark &2 &U$^{2}$ &2.77 &74.90 &10.34\\
			2d &3.76 &\checkmark &3 &U$^{2}$ &2.77 &74.66 &10.41\\
			2e &3.76 &\checkmark &4 &U$^{2}$ &2.80 &75.12 &10.55\\
			2f &3.76 &\checkmark &5 &U$^{2}$ &2.77 &74.74 &10.29\\
			\specialrule{0.1em}{0.25pt}{0.25pt}
			3a &3.48 &$\times$ &3 &U &2.68 &72.93 &10.09\\
			3b &5.70 &$\times$ &3 &Transformer &2.62 &71.92 &9.85\\
			3c &12.85 &$\times$ &3 &Conformer &2.67 &72.93 &10.15\\
			\specialrule{0.1em}{0.25pt}{0.25pt}
	\end{tabular}}
	\caption{Ablation study on WSJ0-SI84 dataset.}
	\label{tbl:ablation-studies}
	\vspace{-0.5cm}
\end{table}
\vspace{-0.2cm}
\subsection{Configuration Details}
\label{configuration-details}
\paragraph{Network configuration.} In the U$^{2}$-encoder and U$^{2}$-decoder, the kernel size and stride of 2D-GLUs are namely set as $\left(1, 3\right)$ and $\left(1, 2\right)$ in the time and frequency axes, and we set the kernel size in the UNet-block as $\left(2, 3\right)$. The number of 2D-conv channels remains 64 by default. Denote the number of (de)encoder layers in the $i$-th UNet-block as $U_{i}$, and then $U = \left\{4, 3, 2, 1, 0\right\}$ where $0$ means no UNet-block is employed. For S-TCN and derivative operators, similar to~{\cite{li2022glance}}, two groups of S-TCMs are utilized, each of which includes four S-TCMs with kernel size and dilation rates of $5$ and $\left\{1, 2, 5, 9\right\}$, respectively. Causal convolution operations are adopted by zero-padding along the past frames.
\paragraph{Training configuration.} We sample all the utterances at 16 kHz. The window size is set as 20 ms, with 50\% overlap between adjacent frames. 320-point FFT is utilized, leading to 161-D in the feature axis. The model is trained on Pytorch platform with a NVIDIA V100 GPU. We use the Adam optimizer $\left(\beta_{1} = 0.9, \beta_{2} = 0.999\right)$ with a batch size of 8 to train the proposed model, and the learning rate is initialized as 5e-4. For WSJ0-SI84, we train the model by 60 epochs, while 30 epochs for DNS-Challenge. The RI loss together with magnitude constraint is utilized as the training loss and the power-spectrum compression strategy is utilized with the compression coefficient empirically set as 0.5~{\cite{li2022glance}}. The learning rate will be halved if the validation loss does not decrease for two consecutive epochs.
\vspace{-0.15cm}
\subsection{Evaluation Metrics}
\label{evaluation-metrics}
Multiple objective metrics are adopted, including narrow-band (NB)~{\cite{rix2001perceptual}} and wide-band (WB) perceptual evaluation speech quality (PESQ)~{\cite{rec2005p}} for speech quality, short-time objective intelligibility (STOI)~{\cite{taal2011algorithm}} and its extended version ESTOI~{\cite{jensen2016algorithm}} for intelligibility, and SISNR~{\cite{le2019sdr}} for speech distortion. For all the metrics, high values indicate better performance.
\renewcommand\arraystretch{1.20}
\begin{table}[t]
	\Huge
	\centering
	\resizebox{\columnwidth}{!}{
		\begin{tabular}{cccccccc}
			\specialrule{0.1em}{0.25pt}{0.25pt}
			\multirow{2}*{Methods} &\multirow{2}*{Do.} &Param. &MACs &\multirow{2}*{RTF} &\multirow{2}*{NB-PESQ$\uparrow$}  &\multirow{2}*{ESTOI$\uparrow$} &\multirow{2}*{SISNR$\uparrow$}\\
			& &(M) &(G/s) & & &(\%) &(dB) \\
			\specialrule{0.1em}{0.25pt}{0.25pt}
			Noisy &- &- &- &- &1.82 &41.97 &0.00\\
			GCRN~{\shortcite{tan2020learning}} &T-F &9.77 &2.42 &0.21 &2.48 &70.68 &9.21 \\
			DCCRN~{\shortcite{hu2020dccrn}} &T-F&\textbf{3.67} &11.13 &0.44 &2.54 &70.58 &9.47 \\
			PHASEN~{\shortcite{yin2020phasen}}  &T-F&8.76 &6.12 &0.37 &2.73 &71.77 &9.38\\
			FullSubNet~{\shortcite{hao2021fullsubnet}} &T-F &5.64 &31.35 &1.35 &2.55 &65.89 &9.16\\
			CTSNet~{\shortcite{li2021two}} &T-F &4.35 &5.57 &0.41 &2.86 &76.15 &10.92 \\
			GaGNet~{\shortcite{li2022glance}} &T-F &5.94 &\textbf{1.63} &\textbf{0.10} &2.86 &76.87 &10.93\\
			Conv-TasNet~{\shortcite{luo2019conv}} &T &4.35 &5.57 &0.73 &2.48  &72.01  &9.99 \\
			DPRNN~{\shortcite{luo2020dual}} &T &4.22 &9.20 &0.77 &2.54 &73.18 &10.21\\
			DDAEC~{\shortcite{pandey2020densely}} &T &4.82 &36.56 &2.40 &2.76 &74.84 &10.85 \\
			DEMUCAS~{\shortcite{defossez2020real}}  &T &18.87 &4.35 &0.28 &2.67 &76.23 &11.08\\
			\textbf{TaylorSENet$^\dagger$(Ours)} &T-F &3.76 &6.14 &0.50 &2.89 &76.95 &10.82 \\
			\textbf{TaylorSENet(Ours)} &T-F &5.40 &6.14 &0.51 &\textbf{2.91} &\textbf{77.76} &\textbf{11.16} \\
			\specialrule{0.1em}{0.25pt}{0.25pt}
	\end{tabular}}
	\caption{Quantitative comparisons with other SOTA systems on WSJ0-SI84 dataset. Scores are averaged upon different testing cases. ``Do.'' denotes the tranform domain of the method.}
	\label{tbl:wsj0-si84-result}
	\vspace{-0.4cm}
\end{table}
\renewcommand\arraystretch{0.85}
\begin{table*}[t]
	\normalsize
	\centering
	\scalebox{0.81}{
		\begin{tabular}{cccccccccc}
			\specialrule{0.1em}{0.25pt}{0.25pt}
			\multirow{2}*{Methods} &\multirow{2}*{Do.}  &\multicolumn{4}{c}{w/ Reverberation} & \multicolumn{4}{c}{w/o Reverberation}\\
			\cmidrule(lr){3-6}\cmidrule(lr){7-10}
			 & &WB-PESQ &NB-PESQ &STOI(\%) &SISNR (dB) &WB-PESQ &NB-PESQ &STOI(\%) &SISNR(dB)\\
			\specialrule{0.1em}{0.25pt}{0.25pt}
			Noisy   &- &1.82 &2.75 &86.62 &9.03 &1.58 &2.45 &91.52 &9.07\\
			NSNet~{\shortcite{reddy2020interspeech}}  &T-F &2.37 &3.08 &90.43 &14.72 &2.15 &2.87  &94.47 &15.61\\
			DTLN~{\shortcite{westhausen2020dual}}   &T-F &- &2.70 &84.68 &10.53 &- &3.04 &94.76  &16.34\\
			DCCRN~{\shortcite{hu2020dccrn}}  &T-F &- &3.32 &- &- &- &3.27 &- &- \\
			FullSubNet~{\shortcite{hao2021fullsubnet}}  &T-F &2.97 &3.47 &92.62 &15.75 &2.78  &3.31 &96.11 &17.29\\
			TRU-Net~{\shortcite{choi2021real}} &T-F &2.74 &3.35 &91.29 &14.87 &2.86  &3.36 &96.32 &17.55\\
			CTS-Net~{\shortcite{li2021two}} &T-F &3.02 &3.47 &92.70 &15.58 &2.94 &3.42  &96.66  &17.99\\
			GaGNet~{\shortcite{li2022glance}} &T-F &3.18 &3.57 &93.22 &16.57 &3.17 &3.56 &97.13 &18.91 \\
			\textbf{TaylorSENet(Ours)} &T-F &\textbf{3.33} &\textbf{3.65} &\textbf{93.99} &\textbf{17.10} &\textbf{3.22} &\textbf{3.59} &\textbf{97.36} &\textbf{19.15}\\
			\specialrule{0.1em}{0.25pt}{0.25pt}
	\end{tabular}}
	\caption{Quantitative comparisons with other state-of-the-art systems on the DNS Challenge dataset. ``-'' denotes no published result.}
	\label{tbl:dns1}
	\vspace{-0.5cm}
\end{table*}
\begin{figure}
	\centering
	\subfigure{
		\begin{minipage}[b]{0.232\linewidth}
			\includegraphics[width=\linewidth]{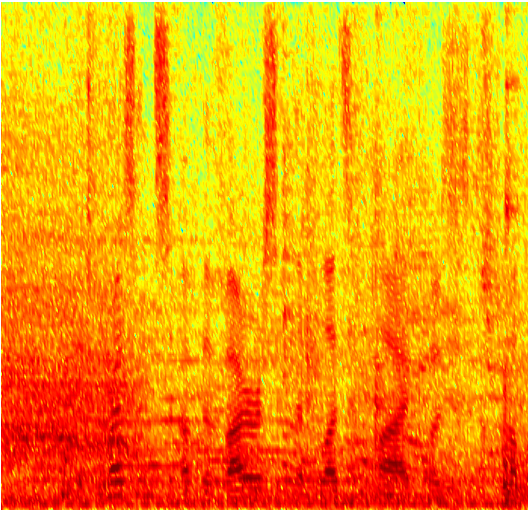}\vspace{-2pt}
			\centering {\scriptsize Mix}\vspace{-4pt}
	\end{minipage}}
	\subfigure{
		\begin{minipage}[b]{0.232\linewidth}
			\includegraphics[width=\linewidth]{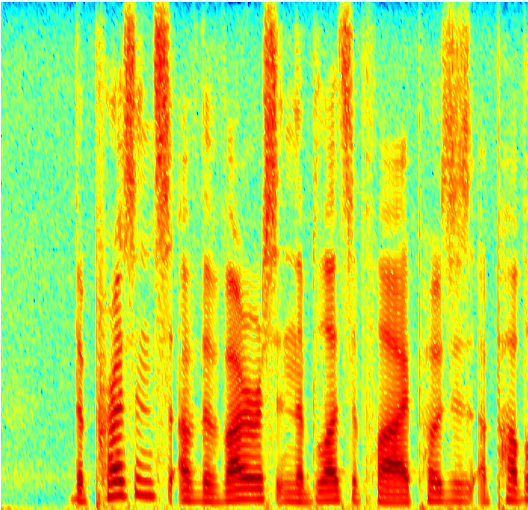}\vspace{-2pt}
			\centering {\scriptsize Clean}\vspace{-4pt}
	\end{minipage}}
	\subfigure{
		\begin{minipage}[b]{0.232\linewidth}
			\includegraphics[width=\linewidth]{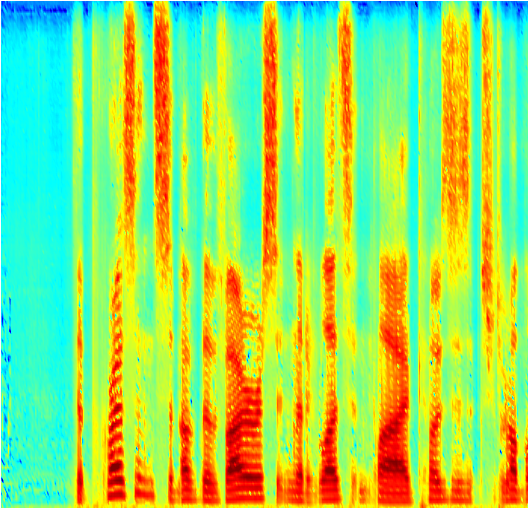}\vspace{-2pt}
			\centering {\scriptsize Esti}\vspace{-4pt}
	\end{minipage}}
	\subfigure{
		\begin{minipage}[b]{0.232\linewidth}
			\includegraphics[width=\linewidth]{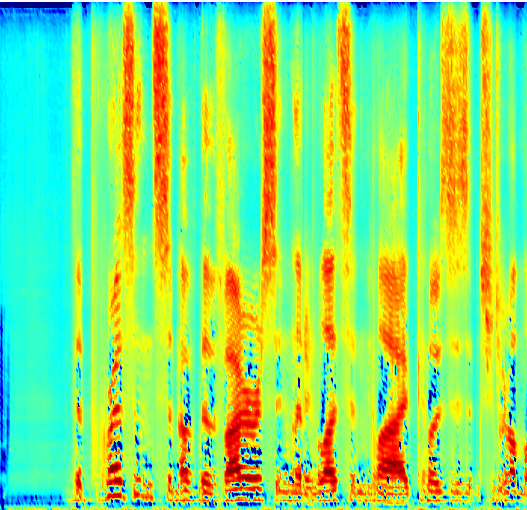}\vspace{-2pt}
			\centering {\scriptsize $\mathcal{F}\left(X\right)$}\vspace{-5.5pt}
	\end{minipage}}
	\subfigure{
		\begin{minipage}[b]{0.232\linewidth}
			\includegraphics[width=\linewidth]{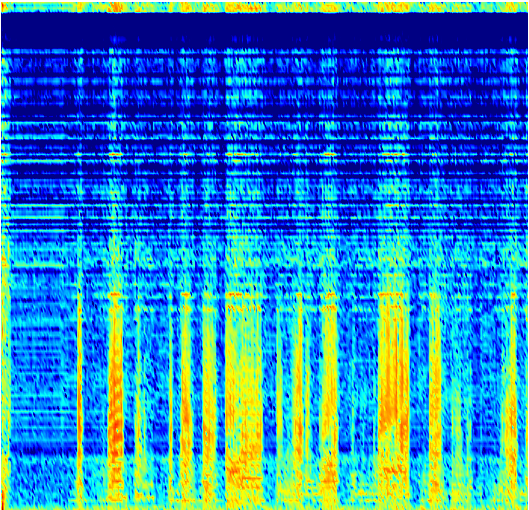}\vspace{-2pt}
			\centering {\scriptsize $\frac{1}{1!}\mathcal{T}\left(1,X,\delta\right)$}\vspace{-4pt}
	\end{minipage}}
	\subfigure{
		\begin{minipage}[b]{0.232\linewidth}
			\includegraphics[width=\linewidth]{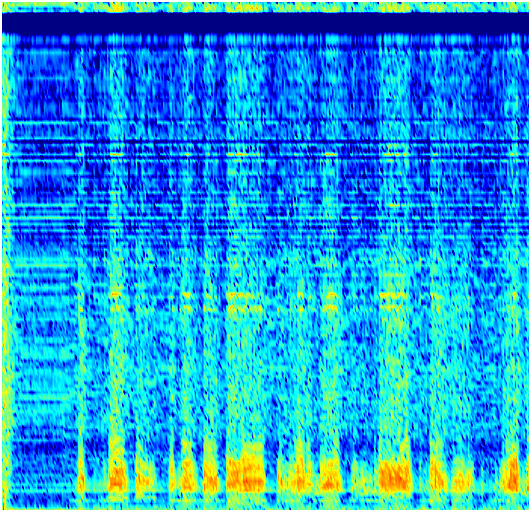}\vspace{-2pt}
			\centering {\scriptsize $\frac{1}{2!}\mathcal{T}\left(2,X,\delta\right)$}\vspace{-4pt}
	\end{minipage}}
	\subfigure{
		\begin{minipage}[b]{0.232\linewidth}
			\includegraphics[width=\linewidth]{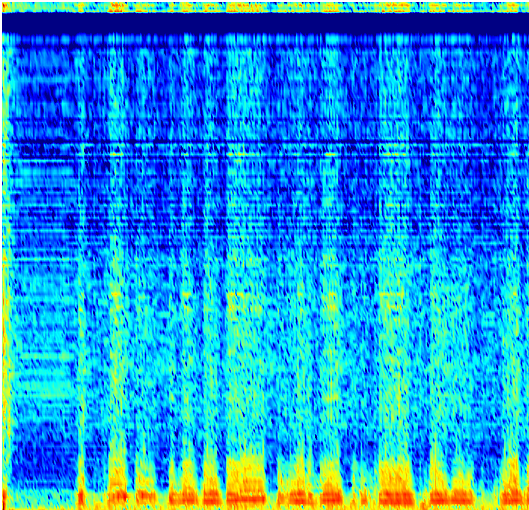}\vspace{-2pt}
			\centering {\scriptsize $\frac{1}{3!}\mathcal{T}\left(3,X,\delta\right)$}\vspace{-4pt}
	\end{minipage}}
	\subfigure{
		\begin{minipage}[b]{0.232\linewidth}
			\includegraphics[width=\linewidth]{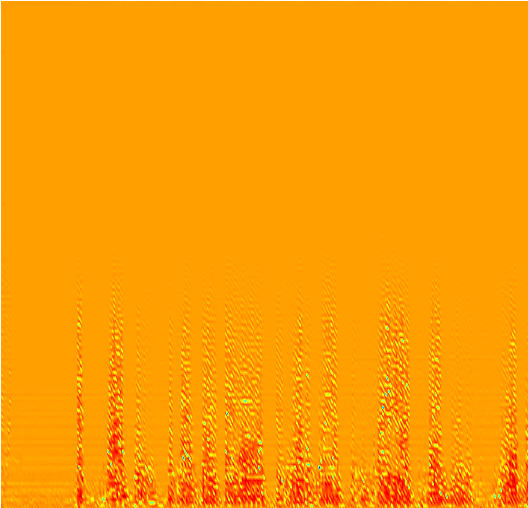}\vspace{-2pt}
			\centering {\scriptsize $\sum_{q}\frac{1}{q!}\mathcal{T}\left(q,X,\delta\right)$}\vspace{-4.5pt}
	\end{minipage}}
	\caption{Estimation visualization of intermediate modules.}
	\vspace{-0.4cm}
	\label{fig:visualization}
\end{figure}
\vspace{-0.2cm}
\subsection{Ablation Study}
\label{sec:ablation-study}
We randomly sample around 100-hour pairs from WSJ0-SI84 corpus to conduct the ablation study, which spans the following three aspects: (1) Is-shared: whether the parameters are shared among multiple high-order modules; (2) $Q$: the number of derivative orders; and (3) Zero-type: network type used in the 0th-order module. We fix the random seed, and PESQ, ESTOI, and SISNR are utilized as the evaluation metric, whose results are shown in Table~{\ref{tbl:ablation-studies}}.
\paragraph{Effect of the parameter-shared scheme for high-order modeling.} In the aforementioned problem formulation, we adopt a trainable network to model the complicated derivative operator for high-order modeling. It is thus necessary to explore whether the parameters can be shared among modules. By comparing the entries from entries 1c-1f to 2c-2f in Table~{\ref{tbl:ablation-studies}}, the non-shared case yields relatively better performance over the shared case, especially in terms of ESTOI and SISNR. We can explain the phenomenon from two aspects. First, the former scheme provides more parameter freedom in the optimization space by independent gradient update, while the latter has to balance all the terms with only one set of parameters. Besides, PESQ mainly focuses on the magnitude envelope similarity while ESTOI and SISNR have a close relation to the phase accuracy~{\cite{wang2021compensation}}, and high-order terms are mainly responsible for phase modification. As a consequence, the nonshared scheme only yields a slightly better PESQ score than the shared case and relatively more notable improvements in terms of ESTOI and SISNR.
\paragraph{Effect of the order number.} As shown in entries 1a-1f of Table~{\ref{fig:framework}}. When $Q = 0$, \emph{i.e.}, no high-order module is employed, it is not surprising to observe the worst performance as only the magnitude is considered and the phase term is neglected. When $Q$ increases from 1 to 3, consistent improvements are achieved among the three metrics, which shows the effectiveness of Taylor series modeling. However, when $Q$ further increases, the performance inclines to saturate and even slightly drop, \emph{e.g.}, $Q = 4$. A similar trend is also observed in entries 2c-2f. This might result from that high-order modules are responsible for complex residual modeling, which has a rather sparse spectral distribution. Therefore, 3-order can be sufficient to approximate the real distribution.

To further validate the mechanism of Taylor-unfolding framework, taking $Q = 3$ as an example, we visualize the output of each order in Figure~{\ref{fig:visualization}}. Noisy, clean, and final estimated spectra are also presented as reference. As we can see, the output from the 0th-order module is quite similar to the clean version, and most noise components are suppressed, which infers that 0th-order indeed serves as the filter to eliminate the noise and captures the overall speech structure in the magnitude domain. Besides, the outputs in the high-order modules seem rather sparse, but we can notice the contour in the harmonic region, indicating that the high-order module effetively captures the sparse residual structure to refine the phase. Remark that as we only supervise the final estimation, despite not strictly following the mathematical definition of Taylor's approximation, the network still learns to allocate the role of 0th-order and high-order terms as expected.

\paragraph{Effect of network types in 0th-order module.} To validate the superiority of the network used in the 0th-order module, we investigate other networks, as shown in entries 3a-3c of Table~{\ref{tbl:ablation-studies}}. ``U'' denotes that no UNet-block is utilized, and ``Transformer'' and ``Conformer'' respectively denote the network is replaced by six transformer and conformer encoding layers, respectively. As we can see, a notable performance drop is observed from entry 1d to 3a, suggesting the effectiveness of UNet-block in feature representation. It is interesting to notice that despite transformer and conformer have shown promising performance in ASR and NLP-related tasks more recently, they are inferior to the proposed U$^{2}$ version herein. This might because they mainly consider the global sequential correlations and ignore the local spectral patterns, which hampers the overall filter estimation. 
\vspace{-0.2cm}
\subsection{Comparison with the State-of-the-Art Methods}
\label{comparison-with-the-state-of-the-art-methods}
\paragraph{WSJ0-SI84.} Configurations of entries 1d and 2d in Table~{\ref{tbl:ablation-studies}} are selected for comparisons with other ten top-performed baselines, where the superscript ``$\dagger$'' denotes that the parameters are shared among high-order modules.  Note that except PHASEN, all baselines are based on causal implementation. The quantitative results are shown in Table~{\ref{tbl:wsj0-si84-result}}, where NB-PESQ, ESTOI, and SISNR are utilized as the evaluation metrics. We also present the number of parameters, multiply-accumulate operations (MACs) per second, and real-time factor (RTF) to evaluate the model computational complexity. As we can see, our method achieves the highest metric scores among all the systems with reasonable trainable parameters and computational complexity. Even with the causal convolutions, our method still dramatically surpasses PHASEN, a noncausal system, which reveals the superiority of our system based on Taylor-unfolding theory. It is interesting to notice that, when parameters are shared, our method will slightly degrade in metric scores, nonetheless, it is also comparable over the state-of-the-art systems but with relatively less trainable parameters.
\paragraph{DNS-Challenge.} To verify the superiority of the proposed SE system in more complicated acoustic scenarios, we report the results on Interspeech2020 DNS-Challenge corpus, as shown in Table~{\ref{tbl:dns1}}. WB-PESQ, NB-PESQ, STOI, and SISNR are used for evaluation. One can see that our method achieves the highest metric performance in both reverberation and anechoic acoustic scenarios. It substantially demonstrates the denoising potential of our method in both reverberant and anechoic environments.
\vspace{-0.35cm}
\section{Conclusion}
We propose a decoupling-style framework based on Taylor's approximation theory for speech enhancement. Specifically, the original complex spectrum reconstruction is decoupled into two parts, namely magnitude estimation and complex residual estimation. For the former, a magnitude filter is devised to suppress the noise components in the magnitude domain. For the latter, multiple trainable modules are unfolded to simulate the complicated derivation operator and estimate the corresponding high-order terms. Afterward, we can recover the target spectrum by superimposition following Taylor's formula. Experiments show that our method achieves state-of-the-art performance over previous top-performed baselines and provides better internal interpretability at the same time.

\bibliographystyle{named}
\small
\bibliography{taylor_ijcai22}

\end{document}